\documentclass[authoryear,preprint,review,12pt]{elsarticle}

\usepackage{amssymb}
\usepackage{physics}

\usepackage{subcaption}
\usepackage{multirow}
\usepackage{hyperref}
\begin{document}

\begin{frontmatter}



% \title{Accelerated Prediction of Fluid Flow in Continuously Stirred Tanks}
\title{Accelerated and data-efficient flow prediction in stirred tanks via physics-informed learning}

% \author{}

% \affiliation{organization={},%Department and Organization

\author[1]{Mahdi Naderibeni\corref{cor1}}
\ead{m.naderibeni@tudelft.nl}
\affiliation[1]{organization={Pattern Recognition and Bio-informatics Group, \\Delft University of Technology},
            % addressline={Van Mourik Broekmanweg 6}, 
            city={Delft},
%          citysep={}, % Uncomment if no comma needed between city and postcode
            % postcode={2628 XE}, 
            country={the Netherlands}}

\author[2]{Liang Wu}
\ead{Liang.wu@dsm-firmenich.com}
\affiliation[2]{organization={Science \& Research and Innovation, dsm-firmenich},
            city={Delft},
%          citysep={}, % Uncomment if no comma needed between city and postcode
            country={the Netherlands}}
\author[1]{David M.J. Tax}%[<options>]
\ead{D.M.J.Tax@tudelft.nl}
\cortext[cor1]{Corresponding author}

\begin{abstract}
The simulation of fluid flows is computationally expensive due to the complexity of its governing partial differential equations. Machine learning models offer a potential surrogate, enabling learning from simulations and significantly faster predictions of flow fields. However, these models require large training datasets, which introduces a trade-off between dataset generation cost and predictive accuracy. In this work, we investigate the relationship between the size of the training-set and accuracy of the prediction when learning steady flow fields in an industrial-scale stirred vessel. A  data set of steady flows is generated using Reynolds Averaged Navier Stokes (RANS) simulations in a range of realistic operating conditions, including impeller speeds and liquid heights. We train implicit neural representations of flow fields and compare purely data-driven and constrained variants. Model performance is evaluated using global mean squared error (MSE), qualitative spatial comparisons of predicted and reference flow fields, and tracer transport simulations. We find that the prediction error decreases monotonically with increasing training data, but also that it exhibits clear diminishing returns beyond moderate dataset sizes. Physics-based constraints significantly improve accuracy and reduce variability across training runs in low-data regimes, and they lead to more stable tracer-transport behavior. Furthermore, reasonable interpolation can be achieved over different impeller speeds and liquid heights. However, these benefits come with an increase in the complexity of training, and their relative advantage diminishes as the training set grows.
\end{abstract}


% \begin{highlights}
% \end{highlights}

\begin{keyword}
Computational Fluid Dynamics \sep Machine Learning \sep Physics-informed Neural Networks \sep Stirred Mixing Tanks

% %% PACS codes here, in the form: \PACS code \sep code

% %% MSC codes here, in the form: \MSC code \sep code
% %% or \MSC[2008] code \sep code (2000 is the default)
% % Fluid dynamics\sep  Machine learning \sep Navier-Stokes equations \sep Physics-informed Neural Network
\end{keyword}

\end{frontmatter}



% \raggedbottom
% \nonpagebreak
% \clearpage

\section{Introduction}

Stirred vessels have widespread applications in chemical and biochemical processes as a result of their favorable performance in mixing and heat transfer \cite{nienow1997mixing}. An essential step in modeling such processes is to simulate the fluid flow by solving the Navier-Stokes equations. Given the tolerable assumptions and accessible compute resources, such simulations are performed at different scales of fidelity. From low-fidelity methods such as Reynolds-averaged Navier-Stokes (RANS) simulations to more computationally intensive approaches such as Large Eddy Simulation (LES) and Direct Numerical Simulation (DNS) \cite{versteeg2007introduction}.

In numerous applications, such as fermentation and crystallization, even the routinely used coarse-grained RANS simulations are computationally prohibitive, requiring researchers and industry practitioners to resort to even lower-fidelity modeling approaches, such as compartment models, despite their inherent limitations \cite{haringa2018industrial}.

In addition, since these simulations involve solving partial differential equations (derived from the conservation laws of mass and momentum, along with transport equations for species and energy), even minor modifications to initial or boundary conditions, such as variations in fluid properties, stirring rate, liquid height, or inlet and outlet flow rates, can significantly alter the solution, necessitating new simulations. Given these limitations and considering that CFD simulations are inherently data-rich, generating extensive spatiotemporal datasets, present a compelling opportunity to leverage data-driven approaches to extract insights and enable predictive modeling \cite{brunton2020machine}.

Recent advances in machine learning have produced a spectrum of approaches to enhance or accelerate CFD simulations. At one end, hybrid methods embed learned components into conventional solvers (e.g. data-driven turbulence closures, surrogate sub-steps, or coarse-to-fine correction operators) to accelerate or improve iterative CFD while retaining the solver's numerical backbone. At the other end, Physics Informed Neural Networks (PINNs) impose the governing PDEs directly on the neural network model training loss and solve for solution maps from coordinates to fields;  even though they still face optimization, scalability, and robustness challenges for large, high Reynolds flows \cite{raissi2019physics}. Neural operator methods (DeepONet \cite{lu2019deeponet}, Fourier Neural Operator \cite{li2020fourier}, etc.) aim to learn mappings between function spaces and provide fast inference for parametric PDE families, though they are typically data-hungry and sensitive to the training distribution. Complementary directions include generative models (conditional GANs, VAEs, diffusion models) that synthesize plausible flow fields conditioned on geometry or boundary/operating parameters \cite{shu2023physics}. Each class of methods trades off data requirements, physical fidelity, inference cost, and ease of coupling with existing CFD workflows. 

In this work, we focus on accelerating the prediction of fluid flow fields for stirred vessels, investigating prediction accuracy as a function of training-data size. We use a CFD solver to generate steady single-phase RANS simulations ($k– \omega$ turbulence model) of an industrial-scale stirred tank across operating conditions parameterized by mixer speed and liquid height, and we train implicit neural surrogates to predict velocity, pressure, turbulent kinetic energy \(k\) and  specific dissipation rate \(\omega\) fields. \\

\textbf{Our contributions are  as follows:}
\begin{itemize}
    \item [-] We generate and publish a controlled dataset of steady flow fields in an industrial stirred vessel using RANS ($k–\omega$) simulations across a sweep of stirring rates and liquid heights.
    
    \item [-] We present a systematic empirical study of the data–accuracy trade-off: evaluating global MSE, local spatial error statistics, and passive-tracer simulations; we show clear diminishing returns in accuracy beyond moderate dataset sizes.
    \item [-] We quantify the benefit of physics constraints: demonstrating that PDE-constrained models substantially improve accuracy, reduce run-to-run variability, and yield more reliable mixing simulations in low-data regimes, while also documenting increased training complexity and the practical challenges of coupling learned fields back into conventional CFD solvers.
\end{itemize}
\section{Data}\label{sec:data_section}

\subsection{Case study}
\begin{figure}[ht]
\centering
  \includegraphics[width=0.5\textwidth]{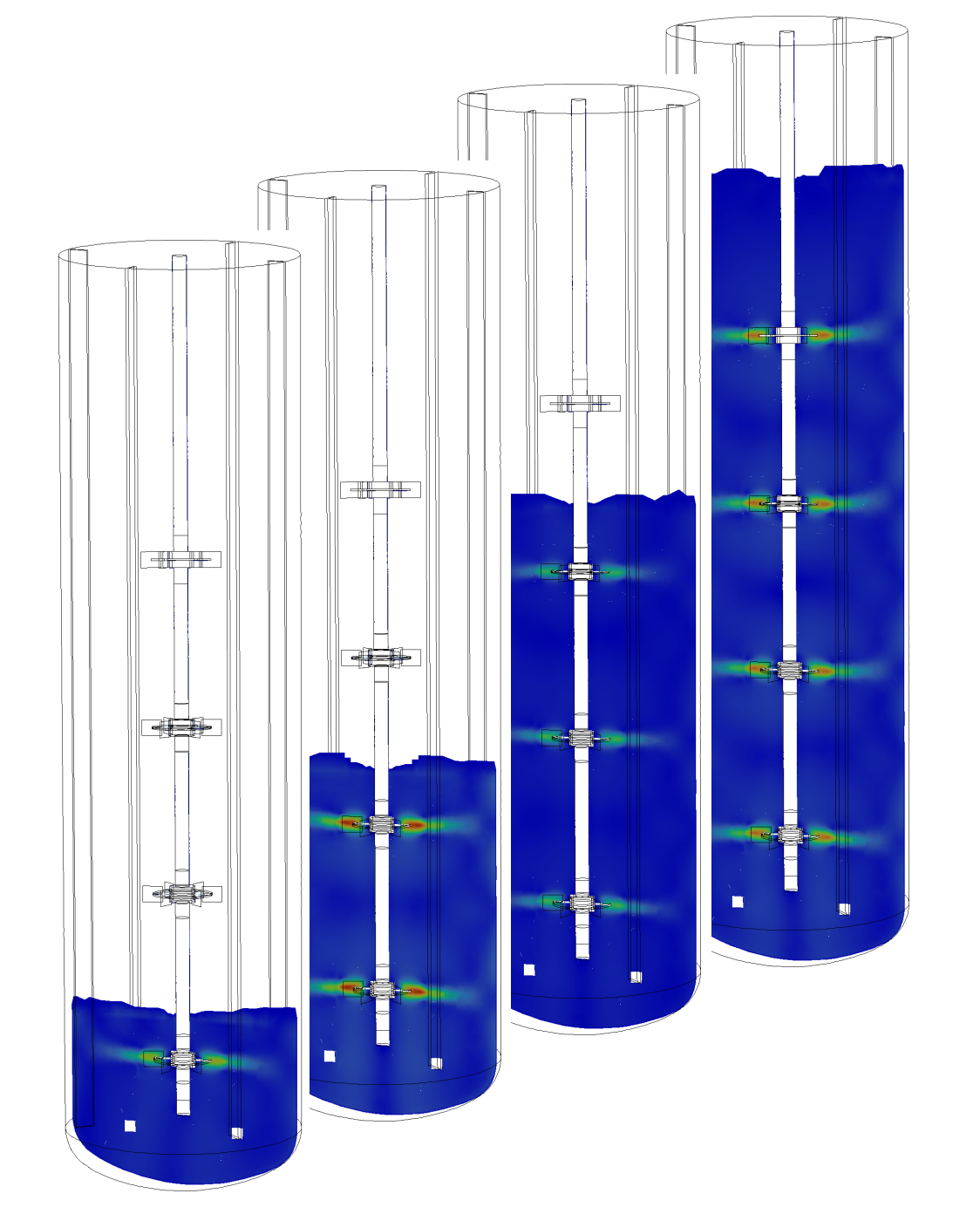} 
    \caption{Tank geometry and four representative velocity magnitude profiles from the dataset, showcasing various liquid heights and stirring rates.}
    \label{fig:geo_data}
\end{figure}

The study considers a $22\,\text{m}^3$ multi-impeller Rushton turbine tank. The vessel has a total height of $8.12\,\text{m}$ and a diameter $T = 2.09\,\text{m}$, and contains four baffles and four Rushton impellers with a diameter of $D = T/3$. To explore the design space, two operational parameters are varied: the liquid height, sampled uniformly between $1.5$ and $6.5\,\text{m}$, and the stirring rate , sampled between $50$ and $150\,\text{rpm}$. A visualization of the vessel geometry together with stacked velocity magnitude profiles from the resulting dataset is shown in Fig.~\ref{fig:geo_data}. Additional technical specifications of the case study can be found in \cite{vrabel2000mixing, haringa2017euler}.

\subsection{Simulation Setup}
Given that the vessel is partially filled, to simulate the two-phase flow in the vessel, we employ the Eulerian multiphase framework coupled with the Shear Stress Transport (SST) $k– \omega$ turbulence model coupled with a mixture turbulence formulation. A production limiter is applied to the turbulence equations to prevent excessive generation of turbulent kinetic energy in stagnation regions. The rotation of the mixer is captured using the Multiple Reference Frame (MRF) approach. Computational resources were provided by the Delft High Performance Computing Center (DHPC) \cite{DHPC2024}, utilizing Ansys Fluent 2024R1 on Intel® Xeon® Gold 6248R processors. Details of the computational mesh and the simulation strategy are provided in \ref{appendix:Simulation-strategy}.

\subsection{Data Pipeline}\label{pipeline}
The simulation results are converted into compact MATLAB (.mat) files for machine learning integration. Each operating condition is represented by 100,000 points sampled uniformly by mesh index, thereby maintaining the non-uniform physical distribution of the original grid (i.e., higher density near the impeller and walls). The data set includes spatial coordinates, velocity, pressure, phase volume fractions, and turbulence closure parameters.

\subsection{Tracer dispersion computations}
To further evaluate the predicted flow fields, we simulated the dispersion of a passive tracer. The tracer was introduced as a patch 0.5 m below the static liquid surface at a radial position of $r=0.7$ m. A numerical probe was positioned at the bottom of the tank ($H=0.5$ m, $r=0.7$ m) to monitor the concentration over time. These simulations were performed on a frozen flow field, where the velocity and turbulence distributions were kept constant while solving the transient species transport equation for a physical duration of 200 s.  This procedure is numerically delicate, and any mismatches in field format, inconsistent discretization or mesh projection, and large local deviations from the reference solution can introduce spurious divergence, violate discrete conservation, or generate non-physical extrema that lead to solver instability. Further numerical details and solver settings are provided in \ref{appendix:mixing-time}.

\section{Related Work}
\textbf{Implicit neural representations.} INRs have emerged as a powerful paradigm for learning continuous functions directly from coordinate inputs. Unlike conventional architectures that operate on discretized data, such as images or voxel grids, INRs model signals as continuous mappings $ f(x) \to y$, where a neural network encodes the target field in its weights. Early successes include applications in computer graphics and vision, such as neural radiance fields (NeRFs) and signed distance functions, which demonstrated that simple multilayer perceptrons can compactly represent complex spatial structures \cite{tancik2020fourier, sitzmann2020implicit, dupont2022data, molaei2023implicit}. 

\textbf{Fourier Feature Maps. } A limitation of early implementations of INRs is their spectral bias (the tendency to learn low-frequency, smooth, slowly varying components of a target function more readily than high-frequency details). This behavior poses challenges in learning complex physical fields, such as turbulent flow or sharp gradient regions, where an accurate reconstruction of fine-scale structures is essential. To mitigate this issue, Fourier feature mappings \cite{tancik2020fourier} and Sinusoidal Representation Networks (SIRENs) \cite{sitzmann2020implicit} were introduced. Fourier feature maps project the input coordinates into a higher-dimensional space spanned by sinusoidal basis functions before passing them into the network, effectively enriching the frequency content of the input and enabling the model to represent high-frequency variations with ease. Similarly, SIRENs employ sinusoidal activation functions to achieve comparable expressivity through architectural design. Both approaches have been shown to significantly improve the convergence rate and accuracy of implicit neural representations and physics-informed networks, particularly in problems involving oscillatory or spatially complex solutions.

\textbf{Physics Informed Neural Networks.} PINNs share a conceptual foundation with INRs as both model continuous functions with coordinate based neural networks. However, while INRs were originally developed in computer vision and graphics to represent shapes, scenes, and images as continuous fields, PINNs emerged within scientific machine learning as a means of embedding physics-based priors into neural function approximators \cite{raissi2019physics}. In both paradigms, PDE constraints can be incorporated into the loss function: in scientific ML, this serves to enforce physical laws and enable forward or inverse PDE solving; in computer vision, similar constraints are introduced to encourage smoothness, learn spatial gradients, or perform image processing tasks \cite{xu2022signal}
\section{Method}
The task we tackle, learning continuous fields governed by partial differential equations (here, the Navier–Stokes equations) differs fundamentally from conventional ML tasks that operate on discrete data arrays. Convolutional networks and attention-based architectures are designed to ingest and process structured grids or tokenized inputs (images, voxels, point clouds) and learn mappings between discrete representations; by construction, they depend on a chosen discretization and often require interpolation when applied across meshes or resolutions. Although discretized formulations using convolutional or graph-based networks could in principle be applied to this problem, such approaches are typically data-intensive and require large, diverse training sets to generalize effectively. In our case, the number of available CFD simulations, each representing a full 3D flow field under specific operating conditions, is limited, making conventional data-hungry architectures unsuitable. Moreover, our objective is to learn multiple continuous fields (e.g., velocity components, pressure, and turbulence quantities) that vary smoothly across the spatial domain. These considerations motivate the adoption of INRs, which efficiently encode continuous functions with relatively few parameters and can be trained directly from scattered point samples without dependence on a fixed grid.

\subsection{Model architecture}\label{architecture}
To parameterize the flow field inside the vessel, we construct a model that learns a continuous mapping from spatial coordinates and operating conditions to the flow variables. The inputs consist of the three-dimensional spatial location and the operating parameters of the vessel, namely the stirring rate and the liquid height. The outputs are the predicted volume fraction, three velocity components, pressure, and turbulence quantities. The model is implemented as a fully connected multilayer perceptron, which serves as an implicit neural representation of the steady-state flow field.

Given that the model input is low-dimensional (3D spatial coordinates), we employ Fourier feature mappings that project the input coordinates into a higher-dimensional space matching the MLP width. The network uses $\tanh$ activation functions, providing smooth and bounded nonlinearities well suited to represent continuous physical fields.

\subsection{Loss Function and Physical Constrains}\label{constraints}
In this work, we consider two model classes: a purely data-driven multilayer perceptron (MLP) trained exclusively on labeled CFD data and physics-constrained variants in which physical knowledge is incorporated into the training objective through PDE residuals following the PINN framework. For both model classes, training mainly aims to minimize the discrepancy between model predictions and the reference CFD solutions ($\mathcal{L}_{\mathrm{data}}$). For this prediction loss, we have:
\begin{equation}
\mathcal{L}_{\mathrm{data}} \;=\; \frac{1}{N}\sum_{i=1}^{N} \big\lVert \hat{\mathbf{y}}(\mathbf{x}_i,\boldsymbol{\mu}_i) - \mathbf{y}_i \big\rVert_2^2,
\label{eq:data_residual}
\end{equation}
where $\{\mathbf{x}_i,\boldsymbol{\mu}_i,\mathbf{y}_i\}_{i=1}^{N}$ denotes the labeled data set (consisting of spatial coordinates, operating condition parameters, and the corresponding target flow variables) and $\hat{\mathbf{y}}$ denotes the predictions of the model. For constrained models, additional loss terms associated with PDE residuals ($\mathcal{L}_{\mathrm{res}}$) are minimized concurrently with prediction loss. This additional loss term penalizes violations of constraints and is constructed as a sum of the error associated with the satisfaction of each constraint; therefore, its specific form depends on the set of equations that are enforced as constraints. This term is defined as the mean squared norm of the combined constraint violations:
\begin{equation} \mathcal{L}_{\mathrm{res}} = \frac{1}{M} \sum_{j=1}^{M} \big\lVert \mathcal{R}(\tilde{\mathbf{x}}_j, \tilde{\boldsymbol{\mu}}_j) \big\rVert_2^2 
\label{eq:pde_residual}
\end{equation}
where the operator $\mathcal{R}$ represents the system of PDEs that are used in the data generation task and $\{\tilde{\mathbf{x}}_j,\tilde{\boldsymbol{\mu}}_j\}_{j=1}^{M}$ denotes the residual points that are arbitrarily sampled across the domain of spatial and operating condition parameters. In this study, we utilize RANS simulations closed with a two equation turbulence model, employing an inertial reference frame formulation within the Multiple Reference Frame (MRF) framework to account for the rotation in the domain (Section \ref{sec:data_section}). Therefore, the continuity, momentum, and transport residuals are given by:
\begin{equation}
\mathcal{R}_{\mathrm{cont}}(\mathbf{x}) = \nabla \cdot \mathbf{u},
\label{eq:continuity}
\end{equation}
\begin{equation}
\mathcal{R}_{\mathrm{u}}(\mathbf{x}) = (\mathbf{u}_{\mathrm{rel}} \cdot \nabla)\mathbf{u} + \boldsymbol{\Omega} \times \mathbf{u} + \frac{1}{\rho} \nabla p - \nabla \cdot \left[ (\nu + \nu_t) \nabla \mathbf{u} \right] - \mathbf{f}
\label{eq:momentum}
\end{equation}
\begin{equation} 
\mathcal{R}_{k}(\mathbf{x}) = (\mathbf{u}_{\text{rel}} \cdot \nabla)k - \nabla \cdot [(\nu + \frac{\nu_{t}}{\sigma_{k}}) \nabla k] - G_{k}+ Y_{k} - S_{k}\end{equation}
\label{eq:kinetic}
\begin{equation} 
\mathcal{R}_{\omega}(\mathbf{x}) = (\mathbf{u}_{\text{rel}} \cdot \nabla) \omega - \nabla \cdot [(\nu + \frac{\nu_{t}}{\sigma_{\omega}}) \nabla \omega] - G_{\omega}+ Y_{\omega} - S_{\omega}
\label{eq:dissipation}
\end{equation}

Here, $\mathbf{u}$ and $p$ denote the predicted absolute velocity and pressure, $\nu$ the molecular kinematic viscosity and $\nu_t$ the eddy viscosity. The relative velocity is defined as $\mathbf{u}_{\mathrm{rel}}=\mathbf{u}-\boldsymbol{\Omega}\times\mathbf{r}$, where $\boldsymbol{\Omega}$ is the imposed angular velocity of the rotating subdomain and $\mathbf{r}$ is the position vector.The term $\mathbf{f}$ denotes additional body forces (e.g. gravity). The variables of turbulence  $k$ and $\omega$ represent the turbulent kinetic energy and the specific dissipation rate, respectively. Details of  SST $\mathrm{k-\omega}$ closure, including production, dissipation, diffusion terms, and model parameters, can be found in the ANSYS Fluent Theory Guide \cite{ansys_fluent_theory}.

\section{Experimental setup}
We consider three closely related model variants for the flow field prediction task. The first is a purely data-driven MLP (denoted MLP) trained exclusively on labeled CFD samples using a supervised MSE loss. The second variant (denoted C-MLP) augments the same architecture and training pipeline with the continuity equation residual (Eq.~\ref{eq:continuity}) in the loss. The third variant (denoted CM-MLP) further incorporates both continuity and momentum conservation residuals (Eq.~\ref{eq:momentum}). Note that while the transport residuals for the turbulent kinetic energy ($\mathcal{R}_k$) and the specific dissipation rate ($\mathcal{R}
_\omega$) are formally defined (see Eqs.~\ref{eq:kinetic} and \ref{eq:dissipation}), they are intentionally excluded from the physics-informed loss formulation in this study. Preliminary experiments including these terms in the loss were conducted, but were found to significantly degrade training stability. In particular, the highly nonlinear source terms and wall-distance-dependent blending functions of the SST $k$–$\omega$ closure introduce substantial stiffness in the automatic differentiation graph, leading to poorly conditioned gradients and hindering optimization. For this reason, the present work focuses on enforcing the fundamental conservation laws, while leaving the incorporation of turbulence-model residuals for future investigation.

\subsection{Data-efficiency study}
We assess data efficiency by varying the number of samples used for training from a total of 72 available samples. The size of the training set spans a low to moderate regime (7, 18, 36, and 54 samples). For each data set size and model variant, we perform multiple independent training runs with different random initializations and data orderings. Performance metrics are reported as aggregated statistics across runs to account for optimization-induced variability.

\subsection{Training protocol}
The input coordinates are passed through Fourier feature mappings prior to the MLP body; the outputs contain flow field variables \((\alpha,u,v,w,p)\). Mini-batches are formed by sampling random (point-index) pairs from training data, preserving the non-uniform spatial sampling induced by the CFD mesh. Residual points for the constrained models were drawn so that they become non-uniform in space but uniform across operating conditions. By sampling the data set by index, the spatial coordinates reflect the original CFD mesh refinement near the walls and rotating parts. Simultaneously, the operating condition variables are drawn uniformly from their specified ranges (Section \ref{sec:data_section}) to provide consistent coverage of the parameter space, independent of local spatial resolution. 

The total loss is implemented as a weighted sum of the residual terms of the data (Eq. \ref{eq:data_residual}) and PDEs (Eq. \ref{eq:pde_residual}), where the total residual PDE loss is scaled by $10^{-3}$ to prevent dominating PDE residuals in the optimization task. In addition,  each PDE residual is divided by its detached magnitude. This normalization yields a scale-invariant contribution, while preserving the gradient direction. As a result, training remains balanced and stable despite variations in residual magnitudes. The training task is performed with the Adam optimizer using an initial learning rate $\eta_0=1\times10^{-3}$ and a StepLR scheduler (step size $=100$ epochs, $\gamma=0.75$) that is advanced once per epoch. Networks are trained for up to 1000 epochs with a mini-batch size of 5000. All experiments are repeated for robustness with $10$ independent random seeds. 

\subsection{Evaluation criteria}
We evaluate the performance of the models using three complementary measures that account for prediction accuracy and physical relevance. First, global precision is quantified by the test mean-squared error (MSE) over all predicted variables \((\alpha,u,v,w,p)\) in the held-out data. We report MSE as a function of training-set size and summarize results across multiple training runs to capture typical performance and variability. Second, the spatial error visualization inspects the local prediction quality by plotting the predicted fields alongside the reference data. While no explicit error metrics are computed, the visual comparison allows us to identify regions where the model deviates from the target flow. Third, tracer-based evaluation assesses the model in a transport phenomena task. Predicted steady velocity fields are used in the CFD solver to simulate passive tracer transport with frozen flow. A probe placed 0.5 m above the tank bottom records the tracer mass fraction over time, and we plot these curves to compare how each model captures the transport dynamics across different training sets and seeds. Together, these experiments provide a clear picture of model behavior: global MSE captures overall fit, spatial errors reveal local shortcomings, and tracer curves show how well the models reproduce practical transport dynamics.

\section{Results}
\subsection{ Global Accuracy}
\begin{figure}[h]
    \centering
    \begin{subfigure}{0.49\linewidth}
    \includegraphics[width=1\textwidth]{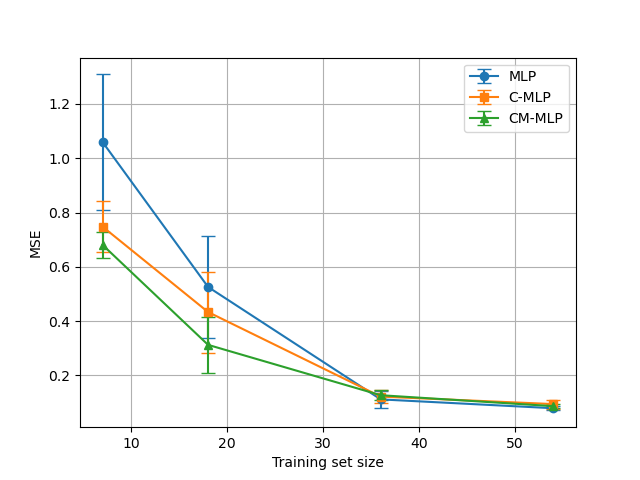}
    \caption{Predictions Mean Squared Error}
    \label{subfig:MSE_vs_trainsize}
    \end{subfigure}
    \begin{subfigure}{0.49\linewidth}
    \includegraphics[width=1\textwidth]{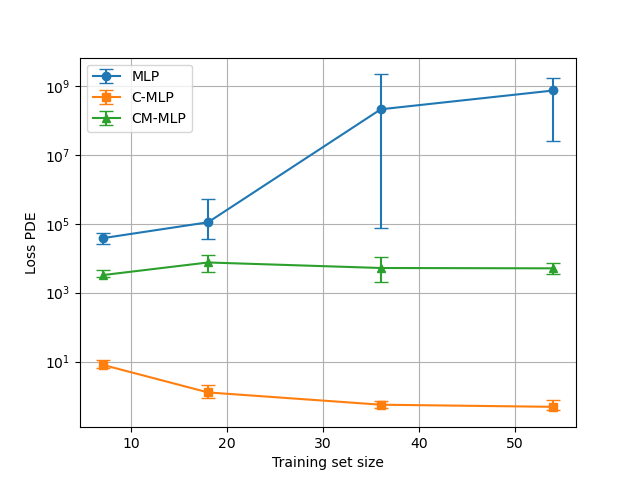}
    \caption{PDE residual Mean Squared Error}
    \label{subfig:LossPDE_vs_train_size}
    \end{subfigure}    
    \caption{Learning curves for MLP, C-MLP, and CM-MLP models. (a) Test prediction error as a function of training set size. (b) Test PDE residual error versus training set size: for MLP and CM-MLP, the reported residual corresponds to the sum of continuity and momentum equations, whereas for C-MLP only the continuity residual is shown.}
    \label{fig:totall_losses}
\end{figure}

We first evaluate models based on global quantitative metrics. Figure \ref{subfig:MSE_vs_trainsize} shows the learning curves (test MSE vs. training-set size). All model classes exhibit monotonically decreasing test MSE as the training data size increases. The constrained models attain substantially lower MSE when only 7 training samples are used. As the size of the training set  grows, the MSE values converge such that all  variants become statistically indistinguishable.

The effect of  the size of the training set on the residuals of the PDEs is shown in Figure \ref{subfig:LossPDE_vs_train_size}. For the C-MLP model, only the continuity residual is shown. As expected, the unconstrained MLP exhibits significantly larger PDE residuals that increase with the size of the  training-set. At first glance, this may seem counterintuitive, since subfigure \ref{subfig:MSE_vs_trainsize} shows that the prediction accuracy improves for all models. However, it is important to note that the PDE residuals here are computed by automatic differentiation of the learned models rather than numerical differentiation of the predicted fields. 

\subsection{ Qualitative Spatial Error Analysis } 
\begin{figure}[h]
    \centering
    \includegraphics[width=1\textwidth]{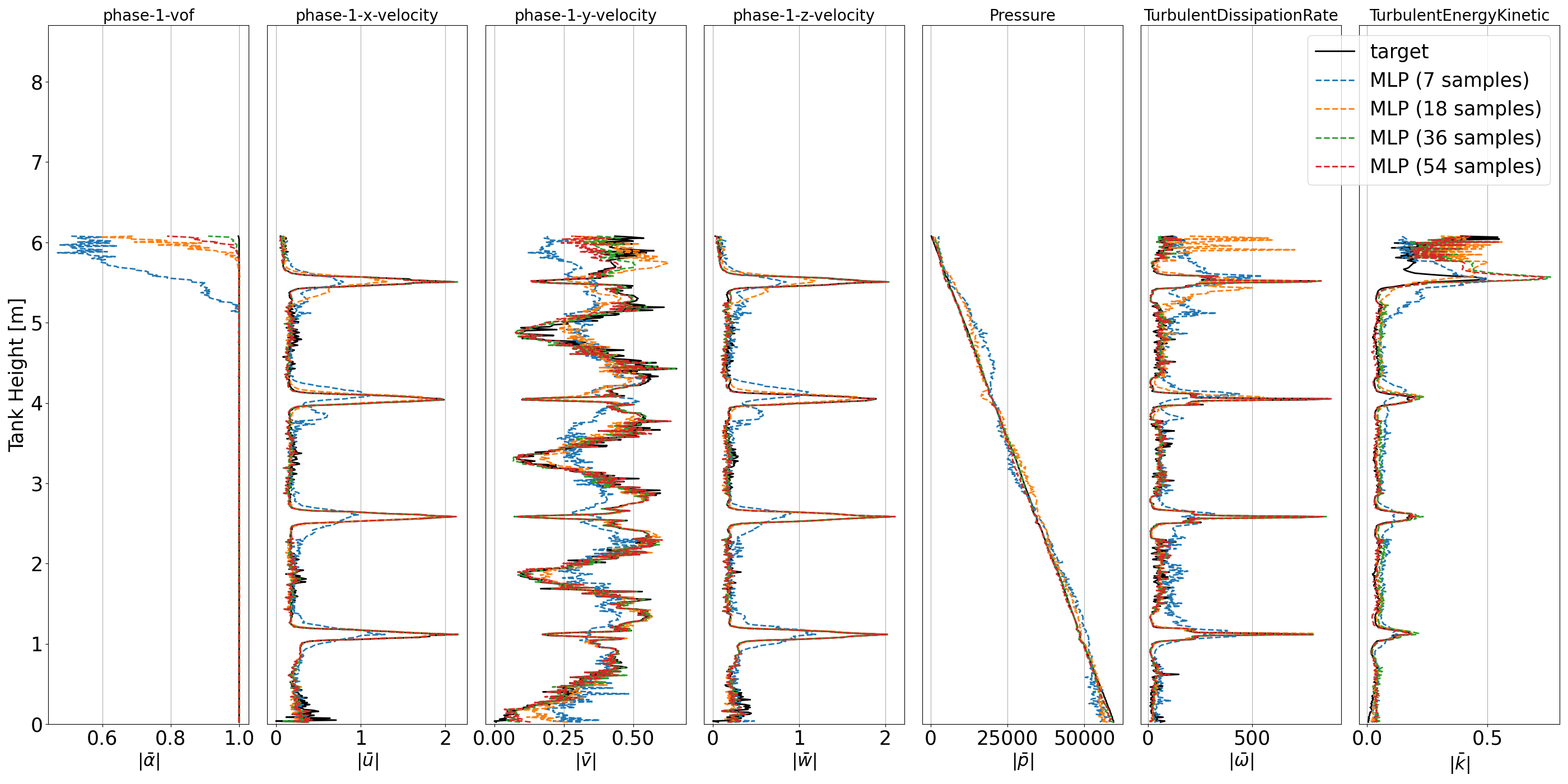}
    \caption{ Predictions made by MLP models (with 10 layers of 100 nodes) trained on different training set sizes:
                [7, 18, 36, 54].}
    \label{fig:100by10_100k_114.526669rpm_6.106744m}
\end{figure}

Although global metrics such as MSE provide a valuable performance summary, they can obscure localized inaccuracies. To perform a more thorough analysis, we examined the spatial distribution of prediction errors for some of the output quantities: the velocity field, pressure, volume fraction, and key turbulence parameters. Figure \ref{fig:100by10_100k_114.526669rpm_6.106744m} shows the predictions for all above mentioned quantities of a test sample by MLP models that are trained on different amounts of training samples. The model trained on only seven samples (blue curve) deviates drastically from the target across the domain. The model trained on 18 samples (orange curve) exhibits noticeable errors primarily near the liquid surface, whereas models trained on larger datasets closely track the target quantities. It is worth noting that as these models are implicit, coordinate-based representations and are trained on labeled and residual points, they encounter substantially more training points at low liquid heights than at higher liquid heights; consequently, prediction errors are systematically smaller in the lower portion of the tank.

\begin{figure}[h]
    \centering
    \includegraphics[width=1\textwidth]{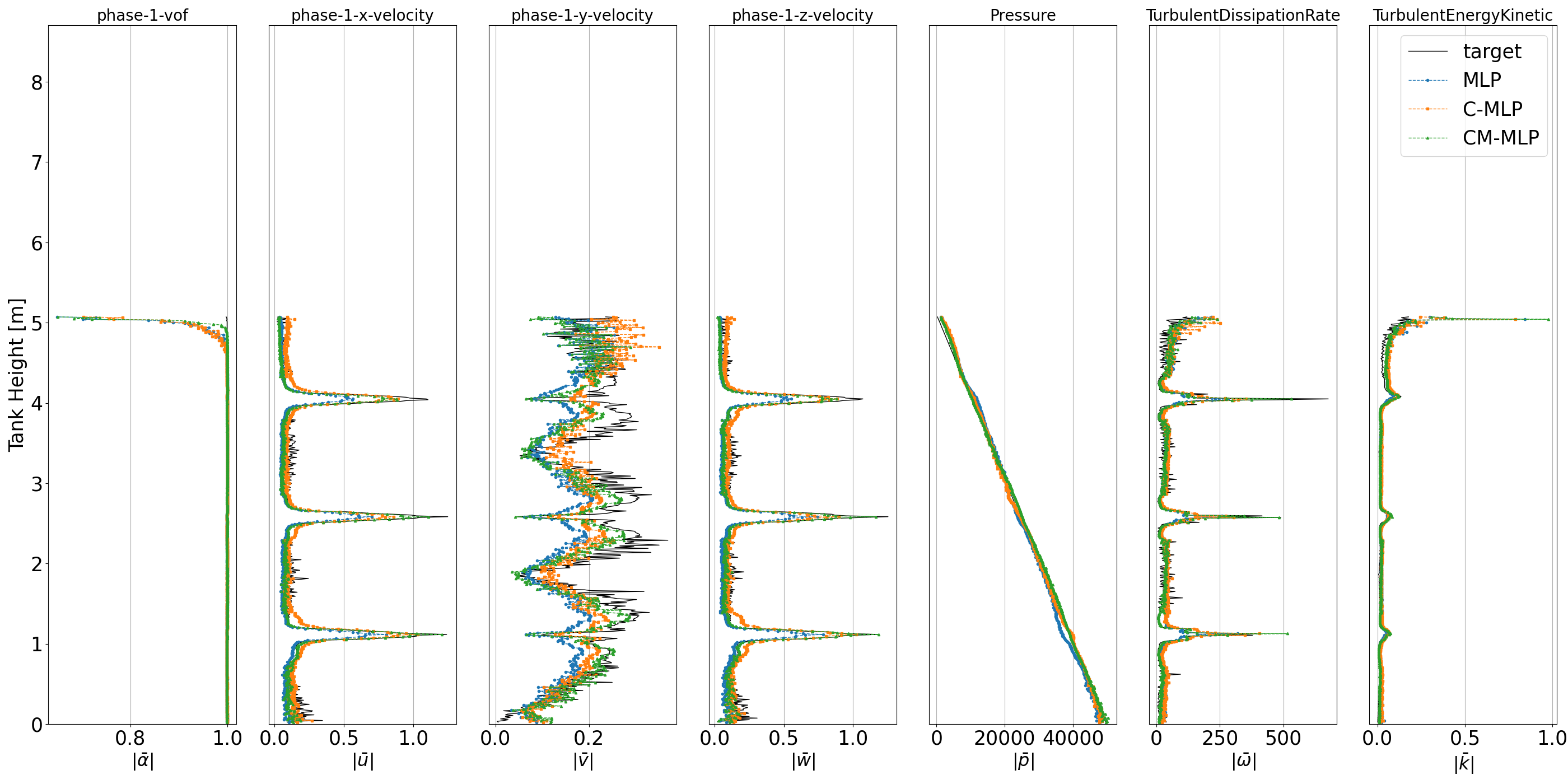}
    \caption{Predictions from the MLP, C-MLP, and CM-MLP models (10 hidden layers, 100 neurons per layer) trained on 18 samples, evaluated at the test operating condition:  stirring rate = 63.6 rpm and liquid height = 5.1 m.}
    \label{fig:63.626111rpm_5.101802m}
\end{figure}

Figure \ref{fig:63.626111rpm_5.101802m} compares the MLP, C-MLP, and CM-MLP models (trained on 18 samples) for a representative test condition. Although all three capture the overall flow structure, the physics-constrained variants, particularly the model enforcing both continuity and momentum equations, exhibit reduced localized errors in regions with strong velocity gradients.

\subsection{Mixing}
\begin{figure}[ht]
    \centering
    \begin{subfigure}{0.49\linewidth}
    \includegraphics[width=1\textwidth]{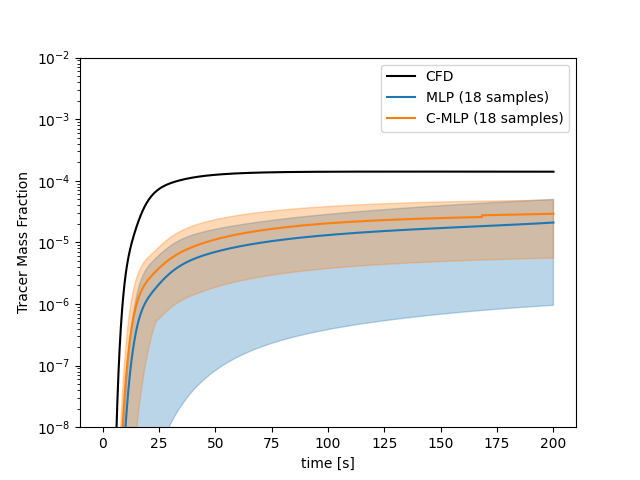}
      \caption{Models trained on 18 samples.}
      \label{subfig:81.184956rpm_3.770235m_0.25}
    \end{subfigure}
    \begin{subfigure}{0.49\linewidth}
    \includegraphics[width=1\textwidth]{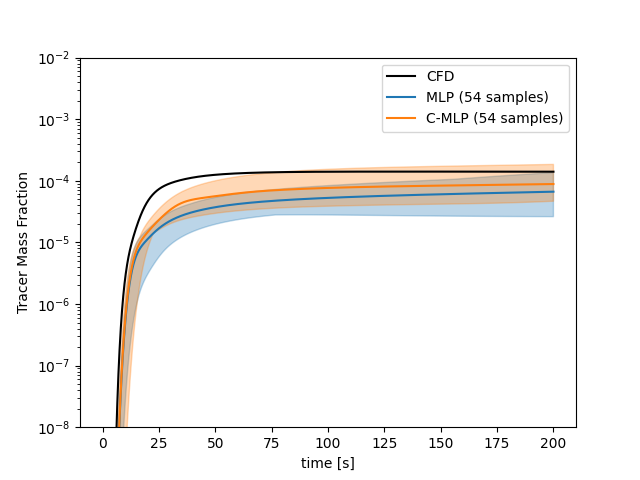}
      \caption{Models trained on 54 samples.}
      \label{subfig:81.184956rpm_3.770235m_0.75}
    \end{subfigure}    
    \caption{Effect of training-set size on tracer concentration predictions at the probe's location. Comparison of MLP and C-MLP models against the CFD reference tested on the sample:  stirring rate =  81.2 rpm, liquid height = 3.8 m.}
    \label{fig:mixing_plots}
\end{figure}

To evaluate the physical consistency of the predictions, we performed tracer transport simulations within the predicted domains. Figure \ref{fig:mixing_plots} compares two model classes evaluated as ensembles of independent runs (different random initialization seeds) for two sizes of the training set . The tracer concentration plots for models trained on 18 samples (Figure \ref{subfig:81.184956rpm_3.770235m_0.25}) exhibit larger inter-run variability and a noticeable end-state bias in the ensemble mean relative to the CFD reference. Both MLP and C-MLP variants show this bias, which decreases with increasing training set size, as expected; however, the bias is consistently lower for C-MLP. Enforcing the continuity constraint alone improves accuracy and reduces inter-run variability, indicating enhanced robustness. Models trained on larger datasets (Fig. \ref{subfig:81.184956rpm_3.770235m_0.75}) show markedly reduced variability and ensemble means closer to the target tracer concentration, producing stable and accurate mixing-time predictions for both model classes; correspondingly, the relative benefit of the constraint diminishes in this data-richer regime. We note that tracer transport simulations for the CM-MLP variant have not been conducted here; nevertheless, we expect a further reduction in both bias and inter-run variability.

% % \clearpage
\section{Conclusions}
We investigated implicit neural surrogates for steady RANS flow fields in an industrial stirred vessel and examined how the size of the training set and PDE-based constraints influence the prediction accuracy and behavior of downstream transport phenomena. Our main observations are summarized below.

\textbf{Data–accuracy trade-off.}
Increasing the number of training operating conditions consistently improves global prediction accuracy. The test MSE decreases monotonically with the size of the dataset for all model classes, and qualitative field comparisons confirm better agreement with the CFD reference as more data are introduced. However, the improvement is not linear: beyond a moderate number of training conditions, the gains become progressively smaller, and the performance of different model variants begins to converge. This suggests that, for this problem, a relatively small set of well chosen operating points, potentially fewer than 20, may already capture the dominant learnable structure, while additional samples primarily contribute to local refinement.

\textbf{Benefit of physics constraints.} Imposing physical constraints during training improves robustness, particularly in data-scarce regimes. Models that enforce continuity, especially those that also enforce momentum conservation, achieve a lower test MSE when only a few training conditions are available. As the data set grows, the gap between constrained and unconstrained models narrows, indicating that sufficient data can partially compensate for the lack of explicit physical regularization. However, constraints provide a clear stabilizing effect when data are limited.

\textbf{PDE residual behavior.}
Evaluation of PDE residuals through automatic differentiation highlights an important distinction between data fit and physical consistency. Although the unconstrained MLP reduces the prediction error with more data, its PDE residuals remain relatively large. In contrast, constrained models maintain lower residual levels by construction. This difference becomes relevant when gradients of the learned fields are used in downstream tasks.

\textbf{Numerical coupling fragility.} Using learned steady fields within a conventional CFD solver for species transport proved to be numerically sensitive. Even when global prediction errors are small, local inconsistencies or discretization mismatches can introduce instability, spurious divergence, or non-physical extrema. Stable coupling required careful interpolation onto the solver mesh, consistency checks, and controlled field introduction. These observations highlight that surrogate accuracy alone does not guarantee seamless integration into established CFD workflows.

\textbf{Outlook.} Several directions could extend this work. More information-efficient sampling strategies can reduce the number of simulations required. Uncertainty quantification would help assess confidence in predictions, particularly in extrapolation regimes. Finally, transfer-learning approaches can improve generalization across operating conditions.
\section*{Declaration of generative AI use}
  The authors employed AI-assisted technologies during the drafting of this manuscript to enhance coherence and linguistic quality. All modified content was subsequently validated by the author, who assume full responsibility for the final published material.

\section*{Acknowledgments}
We gratefully acknowledge the insightful discussions and valuable feedback provided by Professor Marcel Reinders, Professor Henk Noorman, Dr. Cees Haringa, Dr. Jiangtao Lu, and Héctor Maldonado de León. This research was supported by the AI4b.io program, a collaborative initiative between TU Delft and dsm-firmenich, with full funding provided by dsm-firmenich and RVO (Rijksdienst voor Ondernemend Nederland).

\section*{Data availability}
All relevant data generated or analyzed during this study is available at the 4TU.ResearchData repository \cite{mnaderibeni2026data}. The source code developed for this work is publicly available in the GitHub repository \cite{Naderibeni2026code}.

\label{}



%% \bibliographystyle{elsarticle-num} 


% \bibliographystyle{elsarticle-num}
% \bibliographystyle{elsarticle-num-names} 
 \bibliographystyle{elsarticle-harv}

\bibliography{manual}

% \section{Appendix}
\appendix
\section{CFD setup}\label{appendix:Simulation-strategy}
\subsection{The Computational mesh}
The domain was discretized using a high-fidelity unstructured mesh consisting of approximately 2 million cells. The mesh utilizes a hybrid topology with mixed and polygonal elements to provide high resolution in high-gradient regions such as the impeller swept volume and near-wall boundaries.
\subsection{Numerical Schemes and Spatial Discretization}
To maintain a balance between stability and accuracy in the multiphase flow field, a pressure-based solver was utilized with Phase Coupled SIMPLE pressure-velocity coupling. The specific discretization schemes are summarized in table \ref{tab:mixing-time-details}:
\begin{table}[htbp]
    \centering
    \caption{Discretization schemes and solver settings}
    \label{tab:mixing-time-details}
    \begin{tabular}{ll}
    \hline
    \textbf{Variable / Setting} & \textbf{Discretization Scheme} \\
    \hline
    Pressure-Velocity Coupling   & Phase Coupled SIMPLE           \\
    Gradient                     & Least Squares Cell Based       \\
    Pressure                     & Second Order                   \\
    Momentum                     & First Order Upwind             \\
    Volume Fraction              & First Order Upwind             \\
    Turbulent Kinetic Energy ($k$)   & Second Order Upwind            \\
    Specific Dissipation Rate ($\omega$) & Second Order Upwind            \\
    Additional Settings & Warped-Face Gradient Correction \\ 
    \hline
    \end{tabular}
\end{table}
\subsection{Convergence Criteria}
The steady-state simulations were monitored using absolute residual criteria. Convergence was defined when all normalized residuals reached a stable plateau below the $10^{-3}$ threshold. 

% Continuity and Momentum: Absolute criteria of $10^{-3}$ for all velocity components ($u, v, w$) across both phases.Turbulence Quantities: $k$ and $\omega$ residuals were strictly monitored to ensure the closure model reached steady-state.Global Balances: In addition to residual monitors, torque on the impellers was monitored to confirm physical convergence of the flow field.

\subsection{Tracer dispersion simulations}\label{appendix:mixing-time}
Tracer dispersion is modeled by solving the unsteady passive scalar advection–diffusion equation. The scalar transport is computed using either velocity fields obtained from CFD simulations or flow fields predicted by the machine-learning model, enabling a rigorous, one-to-one comparison of the transport dynamics captured by each approach. The numerical setup and discretization parameters used for the transient species transport simulations are summarized in table \ref{tab:species_params}.

\begin{table}[htbp]
    \centering
    \caption{Numerical parameters for the transient species transport simulations.}
    \label{tab:species_params}
    \begin{tabular}{ll}
    \hline
    \textbf{Parameter} & \textbf{Value} \\ 
    \hline   
    Solver Mode & Transient (Frozen Flow) \\
    Total Physical Time & 200 s \\
    Molecular Diffusivity ($D_m$) & $6 \times 10^{-10}$ m$^2$/s \\
    Spatial Discretization & First-Order Upwind \\
    Temporal Discretization & First-Order Implicit \\ 
    Time Step & 0.05 s\\
    \hline
    \end{tabular}
\end{table}

\end{document}